\documentclass[twocolumn,english,aps,prl,amsmath,amssymb,superscriptaddress]{revtex4}
\usepackage[latin9]{inputenc}
\usepackage{color}
\usepackage{amstext}
\usepackage{amssymb}
\usepackage{graphicx}

\makeatletter
\@ifundefined{textcolor}{}
{%
 \definecolor{BLACK}{gray}{0}
 \definecolor{WHITE}{gray}{1}
 \definecolor{RED}{rgb}{1,0,0}
 \definecolor{GREEN}{rgb}{0,1,0}
 \definecolor{BLUE}{rgb}{0,0,1}
 \definecolor{CYAN}{cmyk}{1,0,0,0}
 \definecolor{MAGENTA}{cmyk}{0,1,0,0}
 \definecolor{YELLOW}{cmyk}{0,0,1,0}
}

\usepackage{color}

\newcommand{\beq}{\begin{eqnarray}}
\newcommand{\eeq}{\end{eqnarray}}

\makeatother
\usepackage{babel}

\begin{document}

\title{Interplay between charge order and superconductivity in the kagome metal KV$_3$Sb$_5$}

\author{Feng Du}

\thanks{These authors contributed equally to this work.}

\affiliation{Center for Correlated Matter and Department of Physics, Zhejiang University, Hangzhou 310058, China}
\affiliation  {Zhejiang Province Key Laboratory of Quantum Technology and Device, Department of Physics, Zhejiang University, Hangzhou 310058, China}

\author{Shuaishuai Luo}

\thanks{These authors contributed equally to this work.}

\affiliation{Center for Correlated Matter and Department of Physics, Zhejiang University, Hangzhou 310058, China}
\affiliation  {Zhejiang Province Key Laboratory of Quantum Technology and Device, Department of Physics, Zhejiang University, Hangzhou 310058, China}

\author{Brenden R. Ortiz}

\affiliation{Materials Department and California Nanosystems Institute, University of California Santa Barbara, Santa Barbara, CA, 93106, United States}

\author{Ye Chen}

\affiliation{Center for Correlated Matter and Department of Physics, Zhejiang University, Hangzhou 310058, China}
\affiliation  {Zhejiang Province Key Laboratory of Quantum Technology and Device, Department of Physics, Zhejiang University, Hangzhou 310058, China}

\author{Weiyin Duan}

\affiliation{Center for Correlated Matter and Department of Physics, Zhejiang University, Hangzhou 310058, China}
\affiliation  {Zhejiang Province Key Laboratory of Quantum Technology and Device, Department of Physics, Zhejiang University, Hangzhou 310058, China}

\author{Dongting Zhang}
\affiliation{Center for Correlated Matter and Department of Physics, Zhejiang University, Hangzhou 310058, China}
\affiliation  {Zhejiang Province Key Laboratory of Quantum Technology and Device, Department of Physics, Zhejiang University, Hangzhou 310058, China}

\author{Xin Lu}

\affiliation{Center for Correlated Matter and Department of Physics, Zhejiang University, Hangzhou 310058, China}
\affiliation  {Zhejiang Province Key Laboratory of Quantum Technology and Device, Department of Physics, Zhejiang University, Hangzhou 310058, China}

\author{Stephen D. Wilson}

\affiliation{Materials Department and California Nanosystems Institute, University of California Santa Barbara, Santa Barbara, CA, 93106, United States}

\author{Yu Song}

\email{yusong\_phys@zju.edu.cn}

\affiliation{Center for Correlated Matter and Department of Physics, Zhejiang University, Hangzhou 310058, China}
\affiliation  {Zhejiang Province Key Laboratory of Quantum Technology and Device, Department of Physics, Zhejiang University, Hangzhou 310058, China}

\author{Huiqiu Yuan}

\email{hqyuan@zju.edu.cn}

\selectlanguage{english}%

\affiliation{Center for Correlated Matter and Department of Physics, Zhejiang University, Hangzhou 310058, China}
\affiliation  {Zhejiang Province Key Laboratory of Quantum Technology and Device, Department of Physics, Zhejiang University, Hangzhou 310058, China}
\affiliation  {State Key Laboratory of Silicon Materials, Zhejiang University, Hangzhou 310058, China}

\begin{abstract}

The kagome metal KV$_3$Sb$_5$ hosts charge order, topologically nontrivial Dirac band crossings, and a superconducting ground state with unconventional characteristics, providing an ideal platform to investigate the interplay between different electronic states on the kagome lattice. Here we study the evolution of charge order and superconductivity in KV$_3$Sb$_5$ under hydrostatic pressure using electrical resistivity measurements.
With the application of pressure, the superconducting transition temperature $T_{\rm c}=0.9$~K under ambient pressure quickly increases to 3.1~K at $p=0.4$~GPa, as charge order progressively weakens.
Upon further increasing pressure, signatures of charge order disappears at $p_{\rm c1}\approx0.5$~GPa and $T_{\rm c}$ is gradually suppressed, forming a superconducting dome that terminates at $p\approx10$~GPa. 
Beyond $p\approx10$~GPa, a second superconducting dome emerges with maximum $T_{\rm c}\approx1.0$~K at $p_{\rm c2}\approx22$~GPa, which becomes fully suppressed at $p\approx28$~GPa. The suppression of superconductivity for the second superconducting dome is associated with the appearance of a unique high-pressure phase% below $T^{*}\gtrsim150$~K
, possibly a distinct charge order. 
%Three regimes are identified through behavior of the intrinsic resistivity up to room temperature, corresponding to the two superconducting domes and the high-pressure phase, underscoring possible roles of the electronic structure and electron-phonon interactions in determining the highly unusual phase diagram.
\end{abstract}

\maketitle

%\section{Introduction}

The kagome lattice provides a rich setting to realize exotic states of matter, including quantum spin liquids \cite{Syozi1951,Han2012,Broholm2020}, topologically nontrivial electronic structures \cite{Ye2018,Liu2018,Kang2019} and collective electronic orders \cite{Wang2013,Isakov2006,Guo2009,Kiesel2013,Wen2010}.
Recently, discovery of the two-dimensional kagome metal series $A$V$_3$Sb$_5$ ($A$ = K, Rb, Cs) \cite{Ortiz2019,Ortiz2020,ortiz2020superconductivity,yin2021superconductivity} sparked immense interest, as they exhibit topological band structures, sizable correlation effects, charge order and superconductivity. This series further exhibits a giant anomalous Hall effect in the absence of magnetism \cite{Yang2020,Ortiz2020,kenney2020absence}, which is proposed to result from an unconventional charge order with chiral character \cite{jiang2020discovery,yu2021concurrence}. While the superconducting pairing symmetry remains unclear, multiple superconducting domes were revealed under applied pressure in CsV$_3$Sb$_5$ \cite{zhao2021nodal,chen2021double}, which may result from distinct and possibly unconventional superconducting states.
 %or a unusual  pressure-dependence of the electronic structure or electron-phonon interactions.
%\ys{Moreover, the presence of multiple van Hove singularities in the electronic structure suggests competing electronic instabilities, allowing for sensitive tuning of the electronic ground states.} 
These results make $A$V$_3$Sb$_5$ an ideal platform to investigate the relationship between charge order and superconductivity on the kagome lattice.

\begin{figure}
	\includegraphics[scale=0.4]{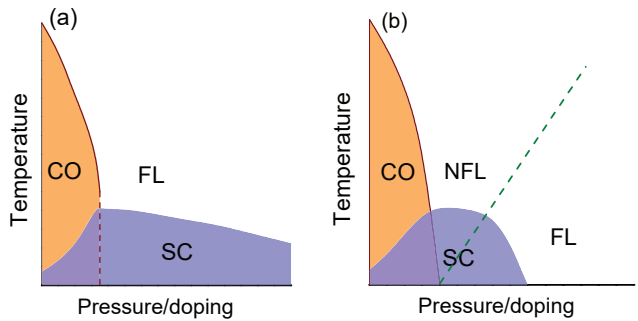} \protect\caption{Schematic phase diagrams of the interplay between charge order and superconductivity (a) where the two orders compete, and (b) when quantum fluctuations associated with charge order are dominant, leading to a fan of non-Fermi-liquid behavior. The orange shaded region corresponds to charge order (CO), and the purple shaded region corresponds to superconductivity (SC). In the normal state without charge order, the system may exhibit Fermi liquid (FL) or non-Fermi-liquid (NFL) behavior.}
	\label{Fig_schematic}
\end{figure}

The interplay between charge order and superconductivity may come in different forms, shown schematically in Fig.~\ref{Fig_schematic}.
Since charge order and superconductivity compete for the same electronic density of states at the Fermi level, the superconducting transition temperature $T_{\rm c}$ is typically enhanced as charge order is suppressed, and evolves more gradually beyond the full suppression of charge order, as depicted in Fig.~\ref{Fig_schematic}(a) %, and when charge order collapses via a first-order quantum phase transition, $T_{\rm c}$ may correspondingly exhibit a jump 
\cite{Gabovich2001,Shen2020,Du2020}.
Alternatively, quantum critical fluctuations associated with charge order may play a dominant role, with a superconducting dome emerging around the quantum critical point, accompanied by a fan of non-Fermi-liquid behavior [Fig.~\ref{Fig_schematic}(b)], similar to unconventional superconductivity around magnetic or nematic quantum critical points \cite{Mathur1998,Yuan2003,Shibauchi2014,Kuo2016}. Although signatures for such an interplay have been suggested \cite{Gruner2017,Klintberg2012,Goh2015}, it remains far from being well-established.
As most studies on the interplay between superconductivity and charge order focused on conventional superconductors, 
it becomes important to examine the corresponding behaviors in systems with unconventional characteristics, such as the $A$V$_3$Sb$_5$ series. In addition, as the $A$V$_3$Sb$_5$ series exhibits multiple Fermi surfaces \cite{Ortiz2020}, it may be susceptible to pressure-induced changes in the electronic structure and electron-phonon interactions, which may promote electronic instabilities distinct from the ambient pressure charge order, allowing for a more nuanced interplay between different order parameters.

In this work, we study the temperature-pressure phase diagram of KV$_{3}$Sb$_5$ single crystals through electrical transport measurements. We find $T_{\rm c}$ increases from $0.9$~K to 3.1~K at 0.4~GPa, with the low-pressure charge order (LPCO) becoming indiscernible above $p_{\rm c1}\approx0.5$~GPa. $T_{\rm c}$ is then gradually suppressed with increasing pressure and terminates at $p\approx10$~GPa, forming a highly asymmetric superconducting dome. A second superconducting dome appears at higher pressures, reaching a maximum $T_{\rm c}\approx1.0$~K at $p_{\rm c2}\approx22$~GPa. Upon further increase of pressure, a high-pressure phase (HPP) distinct from the LPCO appears, evidenced by a hysteretic anomaly in resistivity. Concomitant with appearance of the HPP, superconductivity is suppressed with increasing pressure and disappears above $p\approx28$~GPa. Our observations suggest a nontrivial evolution of the electronic structure and the electron-phonon interaction under pressure, with qualitatively different intrinsic resistivity up to room temperature correlated with distinct ground states.
These findings highlight the $A$V$_3$Sb$_5$ series as a host for tuning between distinct electronic instabilities, and indicate the interplay of superconductivity with the LPCO and the HPP to be mainly driven by their competition, with quantum critical fluctuations playing a minimal role.

%\section{Experimental Details}
Single crystals of KV$_3$Sb$_5$ were grown using a self-flux method, with physical properties under ambient pressure reported previously \cite{ortiz2020superconductivity}. While K-deficiencies may be utilized to achieve unusual transport behaviors \cite{wang2020proximityinduced}, they also significantly increase residual resistivity and suppress superconductivity \cite{Ortiz2019}. Therefore, detailed characterization was performed to ensure that our samples exhibit minimal K-deficiencies \cite{SI}. Electrical resistivity measurements under pressure were carried out using a piston-cylinder cell (PCC) and a diamond anvil cell (DAC), with Daphne oil 7373 or silicon oil as the pressure medium to ensure hydrostaticity of our measurements \cite{SI}.

%\section{Results}

\begin{figure}
	\includegraphics[scale=0.44]{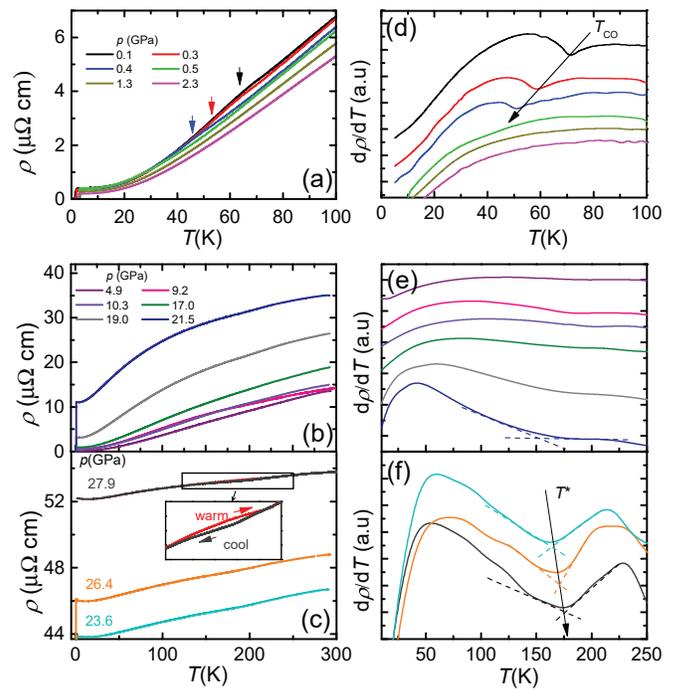} \protect\caption{In-plane resistivity $\rho(T)$ of KV$_3$Sb$_5$ under pressures from (a) 0.1 to 2.3~GPa, (b) 4.9 to 21.5~GPa and (c) 23.6 to 27.9~GPa. The corresponding d$\rho(T)$/d$T$ are respectively shown in (d), (e) and (f). Data in (a) are measured in a piston-cylinder cell, and data in (b) and (c) are measured in a diamond anvil cell. Aside from the one $\rho(T)$ curve at 27.9~GPa which is measured upon warming, all data are taken upon cooling. The inset in (c) zooms into the temperature region with hysteretic resistivity at 27.9~GPa.}
	\label{Fig_normal}
\end{figure}

Measurements of the electrical resistivity $\rho(T)$ with pressures up to 2.3~GPa are shown in Fig.~\ref{Fig_normal}(a), with the corresponding d$\rho$/d$T$ curves shown in Fig.~\ref{Fig_normal}(d). A clear anomaly associated with $T_{\rm CO}$ can be seen at $0.1$~GPa, and as pressure is increased up to 0.4~GPa, the anomaly clearly moves to lower temperatures, before becoming indiscernible at $0.5$~GPa. These results imply that charge order associated with $T_{\rm CO}$ disappears rapidly upon pressure-tuning.
It should be noted that the resistivity anomaly associated with $T_{\rm CO}$ is weaker in KV$_3$Sb$_5$ compared to CsV$_3$Sb$_5$, and it further weakens upon the application of pressure. As the strength of the resistivity anomaly reflects the size of the underlying electronic order parameter, this suggests that in addition to the suppression of $T_{\rm CO}$, the magnitude of the LPCO is also reduced under pressure in KV$_3$Sb$_5$. 

The electrical resistivity $\rho(T)$ in the pressure range 4.9~GPa to 21.5~GPa are shown in Fig.~\ref{Fig_normal}(b), with the corresponding d$\rho$/d$T$ curves shown in Fig.~\ref{Fig_normal}(e). In combination with results in Fig.~\ref{Fig_normal}(d), it can be seen that
d$\rho$/d$T$ does not exhibit clear anomalies from 0.5~GPa to 19~GPa, suggesting no detectable electronic orders that compete with superconductivity in this pressure range. $\rho(T)$ for pressure from 23.6~GPa to 27.9~GPa are shown in Fig.~\ref{Fig_normal}(c), with the corresponding d$\rho$/d$T$ curves shown in Fig.~\ref{Fig_normal}(f). For these pressures, a clear dip is observed in d$\rho$/d$T$, which increases in temperature with increasing pressure. Furthermore, the corresponding anomaly in $\rho(T)$ exhibits a clear hysteresis upon cooling and warming [inset of Fig.~\ref{Fig_normal}(c)], indicating the anomaly to be associated with a HPP appearing via a first-order phase transition.
The HPP's hysteretic nature, significantly higher onset temperature $T^{*}$, and the increase of $T^{*}$ with increasing pressure all indicate the HPP to be distinct from the LPCO. 
Moreover, $\rho(T)$ in the pressure regime with the HPP is qualitatively different from those in Figs.~\ref{Fig_normal}(a) and (b), exhibiting a much reduced $\delta\rho=\rho(300{\rm~K})-\rho_0$ ($\rho_0$ being the resistivity just above the onset of superconductivity). A similar anomaly in d$\rho$/d$T$ is also observed at 21.5~GPa in Fig.~\ref{Fig_normal}(e), pointing to appearance of the HPP already at 21.5~GPa. However, the anomaly is much less prominent at this pressure, and $\rho(T)$ exhibits a larger $\delta\rho$, different from behaviors in Fig.~\ref{Fig_normal}(c). This suggests that appearance of the  HPP is first-order-like upon pressure-tuning, with the HPP partially stabilized over the sample volume at 21.5~GPa. We note that $\rho_0$ appears anomalously large at high pressures, which could be related to structural defects of the sample, possibly caused by structural distortions associated with the HPP and the increased susceptibility to sample fracturing under high pressures.    

\begin{figure}
	\includegraphics[scale=0.41]{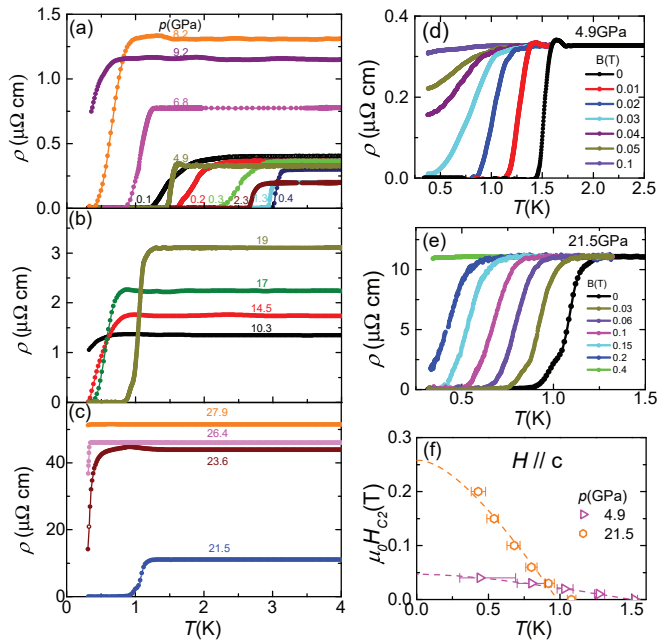} \protect\caption{In-plane resistivity $\rho(T)$ of KV$_3$Sb$_5$ under pressures from (a) 0.1 to 9.2~GPa, (b) 10.3 to 19~GPa and (c) 21.5 to 27.9~GPa, zoomed in for $T\leq4$~K. Low-temperature resistivity $\rho(T)$ under various $c$-axis magnetic fields for (d) 4.9~GPa and (e) 21.5~GPa. (f) The upper critical field of KV$_3$Sb$_5$ as a function of temperature under 4.9~GPa and 21.5~GPa. Fits to the WHH model are shown as dashed lines.}
	\label{Fig_SC}
\end{figure}

Figs.~\ref{Fig_SC}(a)-(c) zooms into $\rho(T)$ for $T\leq4$~K, focusing on the evolution of superconductivity upon pressure-tuning. As can be seen, $T_{\rm c}$ increases with increasing pressure up to $0.4$~GPa, then decreases slowly at higher pressures and becomes strongly suppressed at $9.2$~GPa [Fig.~\ref{Fig_SC}(a)], forming a highly asymmetric superconducting dome with maximal $T_{\rm c}$ near the border of the LPCO. Upon further increase of pressure, $T_{\rm c}$ is first enhanced with increasing pressure from 10.3 to 19~GPa [Fig.~\ref{Fig_SC}(b)], and then decreases from 21.5 to 27.9~GPa [Fig.~\ref{Fig_SC}(c)], forming a second superconducting dome. 

To probe the superconducting state associated with the two superconducting domes, we measured resistivity under an applied magnetic field along the $c$-axis at $4.9$~GPa and $21.5$~GPa, respectively shown in Figs.~\ref{Fig_SC}(d) and (e). The upper critical fields $\mu_0 H_{\rm{c2}}(T)$ are determined as when $\rho(T)$ drops to 
$\rho_0/2$, and are summarized in Fig.~\ref{Fig_SC}(f). $\mu_0 H_{\rm{c2}}(T)$ could be fit with the Werthamer-Helfand-Hohenberg (WHH) model \cite{WHH} for both pressures, with fits shown in Fig.~\ref{Fig_SC}(f). %We find that in addition to the evolution of $T_{\rm c}$ with pressure, $\mu_0 H_{\rm c2}(T=0)$ exhibits an additional nontrivial dependence on pressure 
As can be seen, although $T_{\rm c}\approx1.5$~K at 4.9~GPa is higher than $T_{\rm c}\approx1.0$~K at 21.5~GPa, superconductivity is suppressed much more quickly under an applied field at 4.9~GPa, pointing to a significant pressure-tuning of the superconducting state not captured by $T_{\rm c}$. This point is also highlighted by a nontrivial evolution of the upper critical field around the LPCO \cite{SI}. 

\begin{figure}
	\includegraphics[scale=0.41]{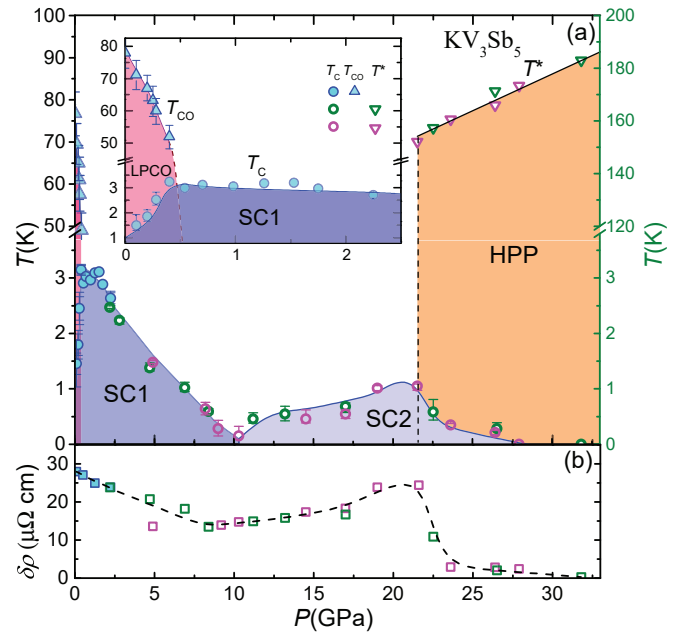} \protect\caption{(a) The temperature-pressure phase diagram of KV$_3$Sb$_5$. Two superconducting regions (SC1 and SC2), the low-pressure charge order (LPCO) and the high-pressure phase (HPP) can be identified in the figure. The inset zooms into the low pressure region, highlighting the interplay between superconductivity and the LPCO. (b) Pressure dependence of the electrical resistivity change $\delta\rho=\rho(300{\rm~K})-\rho_0$. The solid line is a guide-to-the-eye. Full symbols correspond to points obtained using a PCC, and empty symbols correspond to points obtained using a DAC. For measurements using a DAC, two samples were studied \cite{SI} and are presented using different symbols.}
	\label{Fig_phase_diagram}
\end{figure}

%Deviations from the WHH model at low temperatures may be related to multiband effects.
%As can be seen, %superconductivity is gradually suppressed for under pressure, although it survives to a much higher field for $p=0.7$~GPa. 
%the upper critical field as a function of temperature $H_{\rm c2}(T)$ can be fit to a Ginzburg-Landau form Fig.~\ref{Fig_field}(d), with the extracted zero-temperature $H_{\rm c2}$ being 0.28~T, 1.25~T and 0.5~T, for $p=0.1$~GPa, $0.7$~GPa and $2.25$~GPa, respectively. The enhancement of $H_{\rm c2}(T=0)$ with pressure is more significant than the corresponding increase in $T_{\rm c}$ at 0.7~GPa, which only increase by around threefold [Fig.~\ref{Fig_field}(d)].

The phase diagram obtained from electrical resistivity measurements under pressure is shown in Fig.~\ref{Fig_phase_diagram}(a), with $T_{\rm c}$ determined from when $\rho(T)$ drops to $\rho_0/2$, $T_{\rm CO}$ from the d$\rho(T)$/d$T$ dip in Fig.~\ref{Fig_normal}(d), and $T^{*}$ from the d$\rho$/d$T$ dip in Fig.~\ref{Fig_normal}(f). In cases where superconductivity onsets but does not drop to $\rho_0/2$ at the lowest measured temperature, $\rho(T)$ is extrapolated to obtain an estimate of $T_{\rm c}$. The phase diagram reveals that while $T_{\rm c}$ is enhanced from 0.9~K under ambient pressure to 3.1~K at $p_{\rm c1}\approx0.5$~GPa ($4.4$~K/GPa), the decrease above $p_{\rm c1}$ is much more gradual, with the superconducting dome terminating at $p\approx10$~GPa ($\approx-0.30$~K/GPa). 
%Empirically, this contrasts with a scenario involving quantum critical fluctuations and non-Fermi-liquid behavior, where $T_{\rm c}$ usually exhibits clear drops on both sides of the putative quantum critical point. 
Such an asymmetric superconducting dome is empirically different from the case of superconductivity emerging near a quantum critical point.
Combined with the persistence of Fermi-liquid behavior up to at least 17~GPa \cite{SI}, our results suggest that competition between superconductivity and the LPCO is mainly responsible for the maximum in $T_{\rm c}$ at the border of the LPCO. A similar mechanism may also account for the $T_{\rm c}$ maxima at $p_{\rm c2}\approx22$~GPa, given the HPP appears in a clear first-order fashion. Therefore, despite the unusual temperature-pressure phase diagram with the HPP, and unconventional features associated with the LPCO, the interplay between superconductivity with both the LPCO and the HPP seem to be dominated by a competition mechanism [Fig.~\ref{Fig_schematic}(a)], in contrast to a scenario involving quantum criticality [Fig.~\ref{Fig_schematic}(b)].  

On the other hand, the strong suppression of superconductivity at $p\approx10$~GPa likely involves a significant modification to the electronic structure or electron-phonon interactions, leading to a clear kink in the pressure-evolution of $\delta\rho$, shown in Fig.~\ref{Fig_phase_diagram}(b). Since $\delta\rho$ results from electron-electron and electron-phonon scattering and is insensitive to impurity or crystal defects, the kink in $\delta\rho$ at $p\approx10$~GPa may result from a Lifshitz transition \cite{zhang2021pressureinduced,ChenCPL} or a non-monotonic evolution of the electron-phonon interactions, which in turn naturally accounts for distinct characters of the LPCO and the HPP stabilized on the two sides of $p\approx10$~GPa. 
Moreover, $\delta\rho$ reduces sharply near $p_{\rm c2}$ where the HPP is stabilized, consistent with the HPP being a novel state distinct from the LPCO, and supports its first-order appearance with increasing pressure. 

Insensitivity of the HPP to applied magnetic field \cite{SI}, its clear signature in resistivity, and presence of thermal hysteresis suggests the HPP may correspond to a charge ordered state or an altered structural phase. In either case, pressure-tuning of the electronic structure and electron-phonon interactions should play a pivotal role in stabilizing the HPP under pressure, %, in which case our findings suggests the electronic structure and electron-phonon interactions to be strongly modified by pressure, allowing for distinct charge orders to be realized at different pressures. 
%However, the exact nature of the HPP still needs to be determined. %In the case of the HPP being a charge order, since characteristics of charge orders are determined by the electronic structure and electron-phonon interactions, our findings suggests that these are strongly modified with the application of pressure, allowing the HPP to be stabilized.  
%The full suppression of superconductivity at 27.9~GPa suggests the HPP suppresses superconductivity more strongly relative to the LPCO. 
%Given the HPP  appears in a first-order-like fashion, an important question is whether its competition with superconductivity is microscopic or volume-wise. The observation of d$\rho(T)$/d$T$ from 23.6 to 27.9~GPa being similar [Fig.~\ref{Fig_normal}(f)] points to the HPP being fully stabilized in this pressure range, without much change in volume fraction. Combined with the continuous suppression of $T_{\rm c}$ in this pressure range, microscopic competition seems to be favored by our results. 
whose exact nature needs to be clarified by further experiments. The observation of the second superconducting dome having maximal $T_{\rm c}$ around $p_{\rm c2}\approx22$~GPa where the HPP appears, suggests a possible role of the HPP in formation of the second superconducting dome. However, the origin of the second superconductivity dome, whether it exhibits a distinct pairing symmetry relative to the first dome, and exact relationship to the HPP, need to be addressed in future works.   
It is interesting to note that a similar second superconducting dome is also observed in CsV$_3$Sb$_5$ under pressure \cite{zhao2021nodal,ChenCPL}, although superconductivity remains robust up to $100$~GPa, compared to KV$_3$Sb$_5$ in which superconductivity is suppressed around $28$~GPa. The origin of such a significant difference between the two systems calls for further studies, with focus on whether a HPP can be stabilized in CsV$_3$Sb$_5$ under pressure and the role of hydrostaticity in determining the temperature-pressure phase diagram.    
%and whether a HPP that strongly suppresses superconductivity is present in CsV$_3$Sb$_5$ above 100~GPa remains to be clarified. 
%This contrast between CsV$_3$Sb$_5$ and $K$V$_3$Sb$_5$ indicates the HPP is more competitive in the latter, which may be related to its overall lower $T_{\rm c}$ across the temperature-pressure phase diagram, relative to CsV$_3$Sb$_5$. Alternatively, observation of the HPP may be related to the better hydrostaticicty obtained in our experimental approach, and how stability of the HPP may be affected by hydrostaticity of applied pressure should be further investigated. %The full suppression of superconductivity at pressures intermediate between the two superconducting phases is observed in both KV$_3$Sb$_5$ and CsV$_3$Sb$_5$ \cite{zhao2021nodal}, and is likely related to a Lifshitz transition endemic to the $A$V$_3$Sb$_5$ series, and may favor distinct superconducting pairing for the two superconducting phases.  

The temperature-pressure phase diagram we uncover in KV$_3$Sb$_5$ contains two distinct superconducting domes, both with optimal superconductivity near the border of a competing electronic order. Such a behavior is highly unconventional, reminiscent of the heavy fermion superconductor CeCu$_2$Si$_2$ \cite{Yuan2003}, where one superconducting dome is associated with a magnetic quantum critical point and the other with a first-order valence instability. In contrast, the first superconducting dome in KV$_3$Sb$_5$ is associated with an highly unusual charge order, and the second with a new high-pressure phase that is possibly a distinct charge order. Furthermore, compared to the evolution of superconductivity near the border of the LPCO in KV$_3$Sb$_3$, CsV$_3$Sb$_5$ exhibits an additional superconducting dome well inside the LPCO regime \cite{chen2021double}. These rich behaviors in $A$V$_3$Sb$_5$ suggest the presence of multiple electronic instabilities proximate in energy, with the balance between them determined sensitively by the electronic structure, electron-phonon interactions and dimensionality ($c/a=$1.633 in KV$_3$Sb$_5$ and 1.694 in CsV$_3$Sb$_5$), highlighting the kagome lattice as an ideal platform for both realizing and manipulating novel states of quantum matter.

In conclusion, we studied the temperature-pressure phase diagram of the kagome metal KV$_3$Sb$_5$ under hydrostatic pressure, and observed two superconducting domes, with the first exhibiting a competition between superconductivity and the low-pressure charge order, and the second associated with a unique high-pressure phase. Our findings suggest pressure significant modifies the electronic structure and electron-phonon interactions, leading to the nuanced evolution of physical properties and phases under pressure. %The charge orders in KV$_3$Sb$_5$ compete with superconductivity, leading maxima in $T_{\rm c}$ near their border, while Fermi-liquid behavior persist throughout the phase diagram.   
%enhanced as charge order is suppressed for pressure up to $\approx0.4$~GPa. Beyond this pressure, charge order is fully suppressed and superconductivity exhibits a weak decrease with increasing pressure up to 2.25~GPa. Such an extended region with enhanced superconductivity and a lack evidence for non-Fermi-liquid behavior disfavor a major role of quantum fluctuations in determining $T_{\rm c}$. Instead, the enhancement of $T_{\rm c}$ observed below $\approx0.4$~GPa likely results from a weakening of charge order, which competes with superconductivity. 
Despite the nontrivial evolution of superconductivity with pressure, the interplay between charge order and superconductivity is most likely dominated by their competition. Our results evidence two superconducting domes and distinct electronic orders stabilized in the kagome metal KV$_3$Sb$_5$, setting the stage for exploring superconductivity on the kagome lattice in the presence of distinct collective electronic orders, and constraining models that capture charge order and superconductivity in the $A$V$_3$Sb$_5$ series.  
%Relative to the superconductivity transition temperature, the upper critical field exhibits a more nuanced evolution with pressure beyond $p_0$, implying additional subtle effects unrelated to charge order. 
%The interplay between charge order and superconductivity that we observe in KV$_3$Sb$_5$ should be relevant across the $A$V$_3$Sb$_5$ series, and constrains models that capture both orders.

%\section{Acknowledgments}

This work was supported by the National Key R\&D Program of China (No. 2017YFA0303100, No. 2016YFA0300202), the Key R\&D Program of Zhejiang Province, China (2021C01002), the National Natural Science Foundation of China (No. 11974306 and No. 12034017), the Science Challenge Project of China (No. TZ2016004), and the Fundamental Research Funds for the Central Universities of China.  S.D.W. and B.R.O. gratefully acknowledge support via the UC Santa Barbara NSF Quantum Foundry funded via the Q-AMASE-i program under award DMR-1906325.  B.R.O. also acknowledges support from the California NanoSystems Institute through the Elings fellowship program.

%Note: After the submission of our work, we became aware of a study that reveals two domes of superconductivity in KV$_3$Sb$_5$ under pressure \cite{zhu2021doubledome}, consistent with findings in our work. 

\bibliography{bibfile}
\end{document}